# Efficient silicon metasurfaces for visible light

*Zhenpeng Zhou,*[†,‡,∥] *Juntao Li,*[†,‡,∥] *Rongbin Su,*[†,‡] *Beimeng Yao,*[†,‡] *Hanlin Fang,*[†,‡] *Kezheng Li,*[†,‡] *Lidan Zhou,*[†] *Jin Liu,*[†,‡] *Daan Stellinga,*[§] *Christopher P Reardon,*[§] *Thomas F Krauss,*[§,†] *Xuehua Wang\**[†,‡]

[†]State Key Laboratory of Optoelectronic Materials and Technologies, Sun Yat-Sen University, Guangzhou, 510275, China

[‡]School of Physics, Sun Yat-Sen University, Guangzhou, 510275, China

[§]Department of Physics, University of York, York, YO10 5DD, UK

ABSTRACT: Dielectric metasurfaces require high refractive index contrast materials for optimum performance. This requirement imposes a severe restraint; devices have either been demonstrated at wavelengths of 700nm and above using high-index semiconductors such as silicon, or they use lower index dielectric materials such as $TiO_2$ or $Si_3N_4$ and operate in the visible wavelength regime. Here, we show that the high refractive index of silicon can be exploited at wavelengths as short as 532 nm by demonstrating a silicon metasurface with a transmission efficiency of 47% at this wavelength. The metasurface consists of a graded array of silicon posts arranged in a square lattice on a quartz substrate. We show full $2\pi$ phase control and we experimentally demonstrate polarization-independent beam deflection at 532nm wavelength. The crystalline silicon is placed on a quartz substrate by a bespoke layer transfer technique and we note that an efficiency >70% may be achieved for a further optimized structure in the same material. Our results open a new way for realizing efficient metasurfaces based on silicon in the visible wavelength regime.

Metasurfaces are ultrathin optical resonant elements that can manipulate optical wavefronts by modifying the phase, amplitude or polarization of light waves on a subwavelength scale.[1-6] Metasurfaces offer new degrees freedom for controlling light beams on a smaller scale and with higher accuracy than is possible with conventional bulky optical components by controlling the optical path length. Initial demonstrations of metasurfaces involved plasmonic resonances, which, however, are rather lossy and exhibit low efficiency.[7, 8] More recently, all-dielectric metasurfaces have drawn attention because of their high transmission.[9-28] Compared to plasmonic metasurfaces, they offer lower loss, yet they still exhibit Mie resonances for both polarisations at optical frequencies,[13] which has been used to realize perfect reflectors,[14] magnetic mirrors[15] and Huygens surfaces.[17, 18]

Because of its high refractive index and compatibility with CMOS processes, silicon is widely used in all-dielectric metasurfaces devices, such as flat lenses,[19-21] achromatic lenses,[22] vortex generators,[23, 24] holograms[25-27] and nonlinear devices.[28] However, most of the silicon metasurface work is performed in the near-infrared wavelength regime and not in the visible. This is because most researchers use amorphous silicon (a-silicon) or polycrystalline silicon (poly-silicon) due to the ease of deposition onto transparent substrates such as glass. The problem with deposited silicon is its high absorption loss in the visible regime. Thin film crystalline silicon (c-silicon) offers a solution to this problem because of its much lower absorption at λ > 500 nm.[13, 29] For example, the single-pass absorption of a 200 nm thin film of c-silicon and a-silicon is around 10.4% and 51.5% at the wavelength of 500nm, respectively. The corresponding refractive index and extinction coefficient of c-silicon and a-silicon are $n_{c-si}$ = 4.142, $n_{a-si}$ = 4.496, $k_{c-si}$ = 0.0324, $k_{a-si}$ = 0.4552. Alternatively, Titanium dioxide ($TiO_2$),[30, 31] silicon nitride ($Si_3N_4$)[32] and silica ($SiO_2$)[33] have been used for all-dielectric metasurfaces based on their high transparency throughout the visible spectrum. For example, a 600 nm thin $TiO_2$ film has recently been patterned into nanofins in order to form a high numerical aperture (NA) metalens for operation at visible wavelengths.[31] Due to the lower refractive index contrast compared to silicon, however, these $TiO_2$ nanofins require very high aspect ratios of 10-15, which makes the fabrication very challenging.

Here, we propose the use of thin film c-silicon as a metasurface material for visible light operation and demonstrate high efficiency polarization-independent operation in transmission at 532 nm wavelength. This demonstration is enabled by our layer-transfer technique, whereby we transfer a 220 nm c-silicon device layer from a Silicon on Insulator (SOI) wafer to a transparent quartz substrate. The maximum aspect ratio of our metasurfaces is 2.47 in the experiment, which makes it easier to fabricate than comparable devices based on $Si_3N_4$ or $TiO_2$. We believe that this method will open a new way to extend the functionalities of metasurfaces efficiently into the visible light regime.

To illustrate the capability of our c-silicon metasurface in transmission and its full $2\pi$ phase control, we consider light propagating through an array of circular c-silicon posts on a subwavelength square lattice (Figure 1).[11, 20] Due to the circular symmetry of the circular posts, our metasurfaces are polarization-independent.

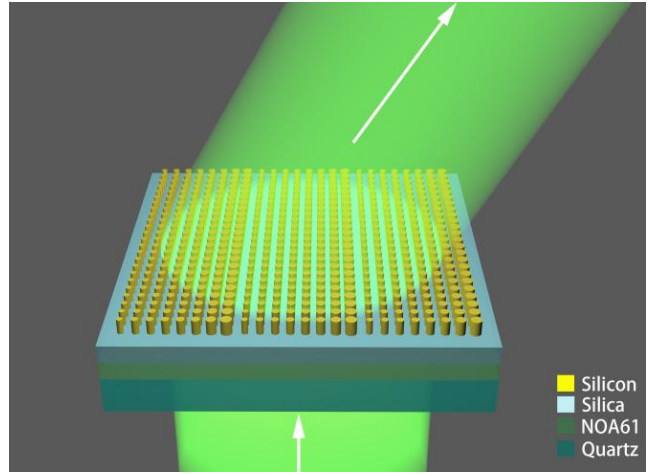

**Figure 1.** Schematic of a gradient metasurface which acts as a beam deflector

We performed the numerical calculation using the rigorous coupled-wave analysis (RCWA) method[34] and analyzed the transmission coefficient and phase of the periodic c-silicon posts by varying the period $a$ from 160 nm to 250 nm and the diameter from $0.2a$ to $0.8a$ at the wavelength 532 nm (Figure 2a, b). In the calculation, the post height h is fixed to be 220 nm, the thickness of the silica film (refractive index $n_{Silica} = 1.45$) and the adhesive NOA61 (Norland Products, Inc.) ($n_{NOA} = 1.55$) underneath are 1 μm, and the refractive index of the quartz substrate is 1.45. As shown in Figure 2, arrays of posts with 190 nm period can achieve large transmission amplitudes while spanning the full range of phases from 0 to $2\pi$ by varying the diameter of the posts from 38 nm to 152 nm. In Table 1, which is based on Figure 2c, we choose eight different diameter posts with $\pi/4$ increments to cover the full 0 to $2\pi$ phase range.

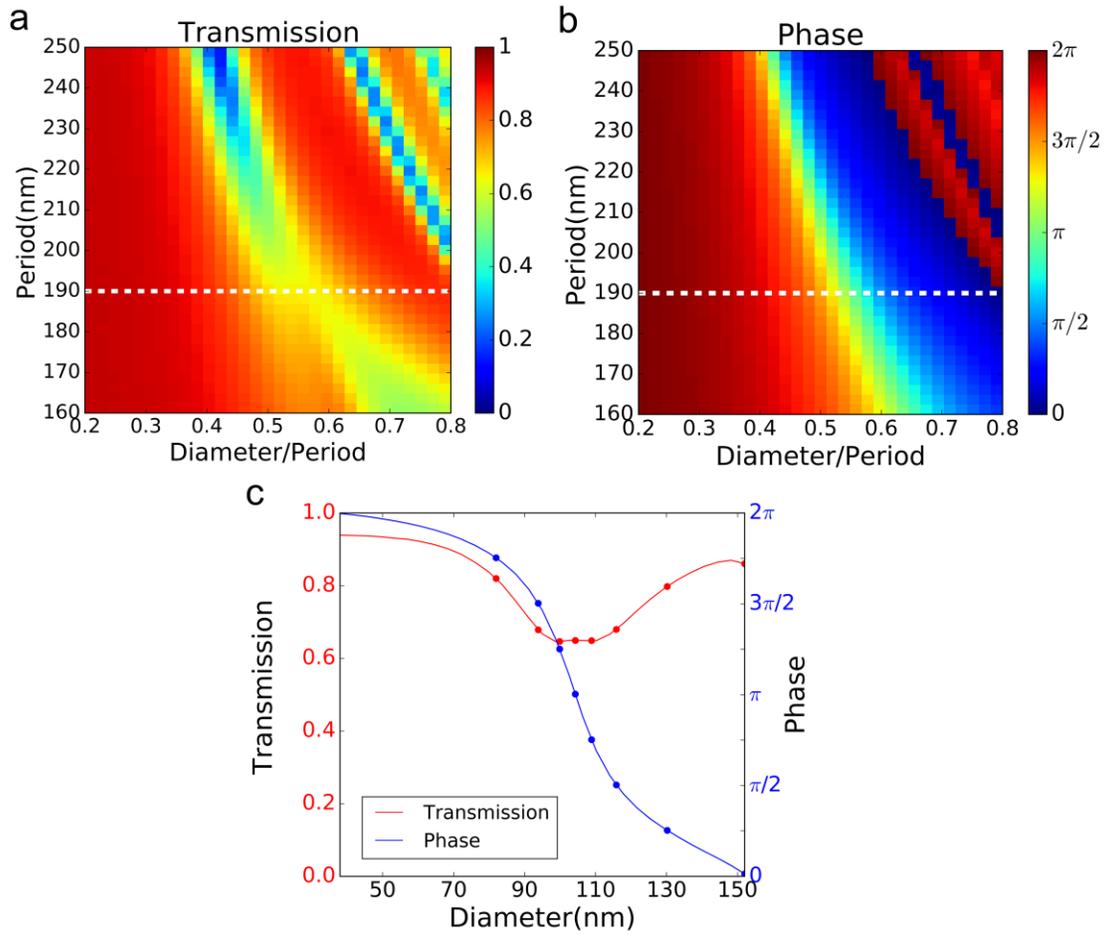

**Figure 2.** Calculation of (a) the transmission and (b) the phase of the periodic c-silicon posts on a square lattice with different periods and diameters. (c) Transmission and phase of the periodic c-silicon posts with 190 nm period for different diameters.

**Table 1.** Diameters of posts with 190 nm period and 220 nm height required to achieve full $2\pi$ coverage in $\pi/4$ steps.

| Phase (rad)   | 0   | $\pi/4$ | $\pi/2$ | $3\pi/4$ | $\pi$ | $5\pi/4$ | $3\pi/2$ | $7\pi/4$ |
|---------------|-----|---------|---------|----------|-------|----------|----------|----------|
| Diameter (nm) | 152 | 130     | 116     | 109      | 104   | 100      | 94       | 82       |

To validate the phase control effect of our c-silicon metasurface, we designed a prism-like refractive index gradient as a beam deflector using the eight phase elements shown in Table 1. The diffraction angle $\theta_t$ of such a gradient surface can be calculated via the generalized Snell's law,[8]

$$n_t \sin \theta_t - n_i \sin \theta_i = \frac{\lambda}{2\pi} \frac{d\Phi}{dx} \qquad (1)$$

where $n_t$ and $n_i$ are the refractive index of the surrounding medium on the transmitted and incident sides, $\theta_i$ is the incident light angle, $\lambda$ is the vacuum wavelength and $d\Phi/dx$ is the phase gradient. In our case, $d\Phi$ equals $\pi/4$ and $dx$ equals the period of 190 nm. Hence we expect that the gradient metasurface will deflect the transmitted beam at an angle of 20.48° to normal incidence.

We first performed an FDTD simulation of the gradient metasurface. As shown in Figure 3a, the refraction angle observed from the phase profile is 20.48°. The same angle can also be calculated from Figure 3b by [35]

$$\theta = \sin^{-1}(k_x/k_0) = \sin^{-1}(0.35) = 20.48° \qquad (2)$$

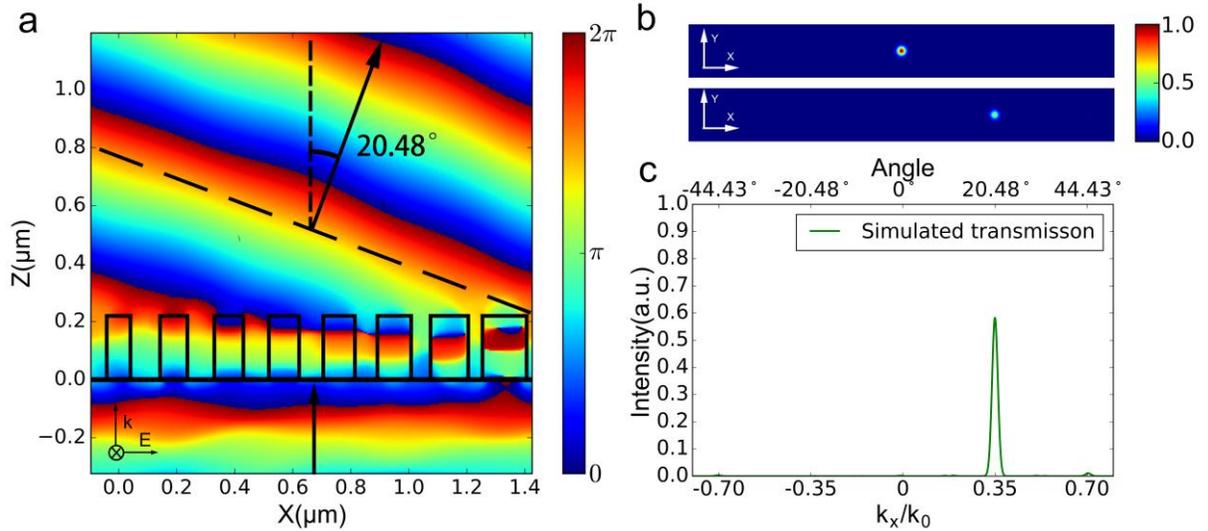

**Figure 3.** FDTD simulation of the gradient metasurface. (a) Phase profile obtained by the metasurface resulting in an angle of deflection of 20.48°. (b) Far-field profiles of the incident light

intensity (top) and transmission intensity (bottom). (c) Transmitted deflected beam intensity normalized to the input signal in $k_x$ direction.

Further, we can calculate the deflection efficiency from Figure 3c by [12]

$$\eta = I_{1st}/I_{input} = 59\% \tag{3}$$

where $I_{1st}$ is the deflecting light intensity in transmission after the metasurface and $I_{input}$ is the incident light intensity. The difference between unity and the observed 59% is mainly caused by the absorption of c-silicon (30%), interface reflectivity (8%), and other diffraction orders (3%). For the 220 nm film used here, the aspect ratio of the fabricated device is 2.7. It is interesting to note that if we were to use a slightly thicker film of 250 nm and a 3.4 aspect ratio, we could increase the transmission further to 71% (Figure 4), which is close to the results obtained with $TiO_2$[30, 31] and which requires a much higher aspect ratio and hence more demanding fabrication.

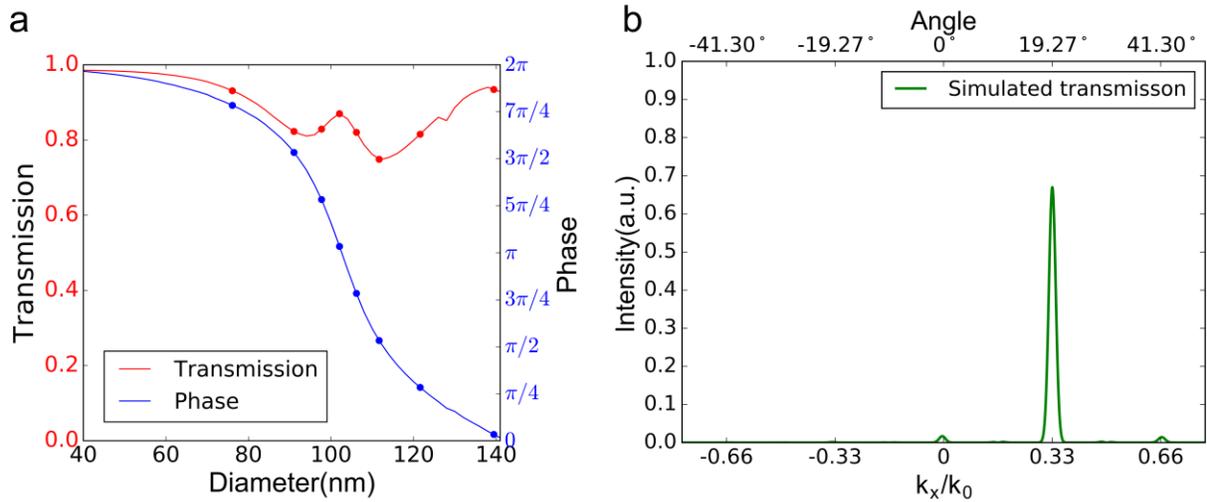

**Figure 4.** Calculation of (a) the transmission and phase of the periodic c-silicon posts with 200 nm period for 250 nm film thickness. (b) FDTD simulation of the transmitted beam intensity normalized to the input signal in $k_x$ direction for the 250 nm thick film.

Thin film c-silicon from a SOI wafer can be transferred to a rigid or a flexible substrate by the lift-off and stamp printing process[36, 37] or by adhesive wafer bonding and deep reactive ion etching (DRIE).[38, 39] We used the latter method because we found it easier to maintain the integrity of the nanostructure. We then used electron beam lithography (EBL) to define the pattern. Our fabrication process is illustrated in Figure 5.

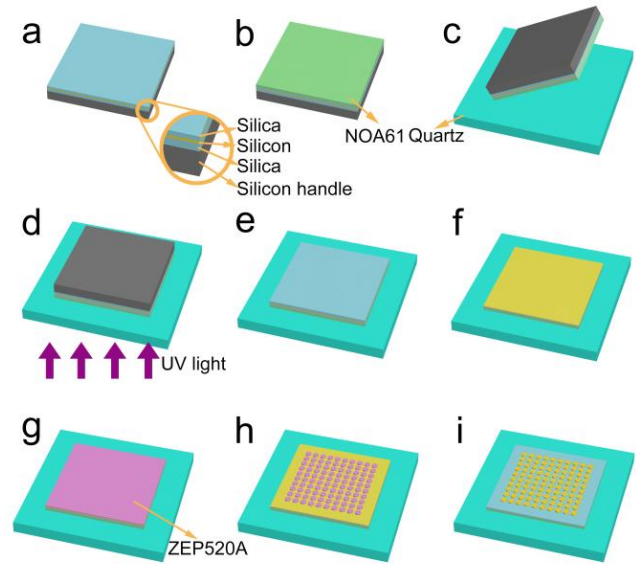

**Figure 5.** Schematic illustration of the c-silicon transfer process and sample fabrication. (a) Deposition of silica on SOI wafer using ICP-CVD. (b) Spin-coating adhesive NOA61 (c) Bonding SOI with fused quartz. (d) Exposing with UV light for 2 hours, followed by baking for 2 days at 50℃. (e) Polishing the silicon substrate to ~ 40 um then removing the remaining silicon substrate by RDIE. (f) Removing silica layer with HF acid. (g) Spinning ZEP520A and depositing Al. (h) Exposing pattern by EBL and removing Al. (i) Transferring pattern to the silicon by ICP then removing the resist by 1165 remover and O2 plasma ashing.

The transfer process is shown in Figure 5a-f. First, we deposit 1 μm silica on a SOITEC SOI wafer comprising a 220 nm thin film c-Silicon layer on 2 μm of silica. This 1 μm silica layer protects the c-silicon from the adhesives and quartz (Figure 5a). Next, we spin the UV light curable adhesive NOA61 on the sample followed by bonding it to the quartz substrate (Figure 5b and 5c). Then the sample is illuminated by 365nm ultraviolet LED light to crosslink the NOA61 polymer for 2 hours. In order to obtain optimum adhesion, the sample is baked at 50℃ for 2 days (Figure 5d). The silicon substrate is then removed by first milling down to near 40 μm followed by DRIE (Figure 5e). Finally, the c-silicon on quartz substrate is obtained by removing the silica of the SOI wafer using HF acid.

The fabrication process of the metasurfaces on the c-silicon by EBL is shown in Figure 5g-i. The sample is spin-coated with 180nm ZEP520A electron beam resist followed by a 50nm aluminum layer (thermal evaporation) to serve as the charge dissipation layer. The pattern is then exposed using a Raith Vistec EBPG-5000plusES electron beam writer at 100keV. After exposure, the aluminum layer is removed by tetramethylammonium hydroxide (TMAH) and the resist is developed with Xylene. Then the pattern transfer is etched using an Oxford Instruments Inductively Coupled Plasma tool.

The overall area of the fabricated gradient metasurface is 200μm x 200 μm. As shown in Figure 6a, we used a 532nm cw laser for illumination, linear polarizers (LP) and a half-wave plate to change the polarization direction of the input light. A 4X objective (Obj1, 0.1 NA) was used to focus the light onto the sample with a spot diameter of ~ 150μm. A 100X objective (Obj2, 0.9 NA) was used to collect the transmitted signal. The real-space and k-space (diffraction order) image of the sample was captured by the CCD, respectively (Figure 6a). From Figure 6b and 6d, we can see that the metasurface directs the light almost entirely into the +1 order, while the other diffraction orders are too weak to be captured by the CCD. The deflection angle is measured to be 21°, which is close to the theoretical calculation and the numerical simulation. The deflection efficiency is calculated to be 47% by using equation (3), which is similar to the transmission measured by the optical power meter. As predicted by theory, the measured deflection efficiency was polarization-independent with only 5% variation (Figure 6c).

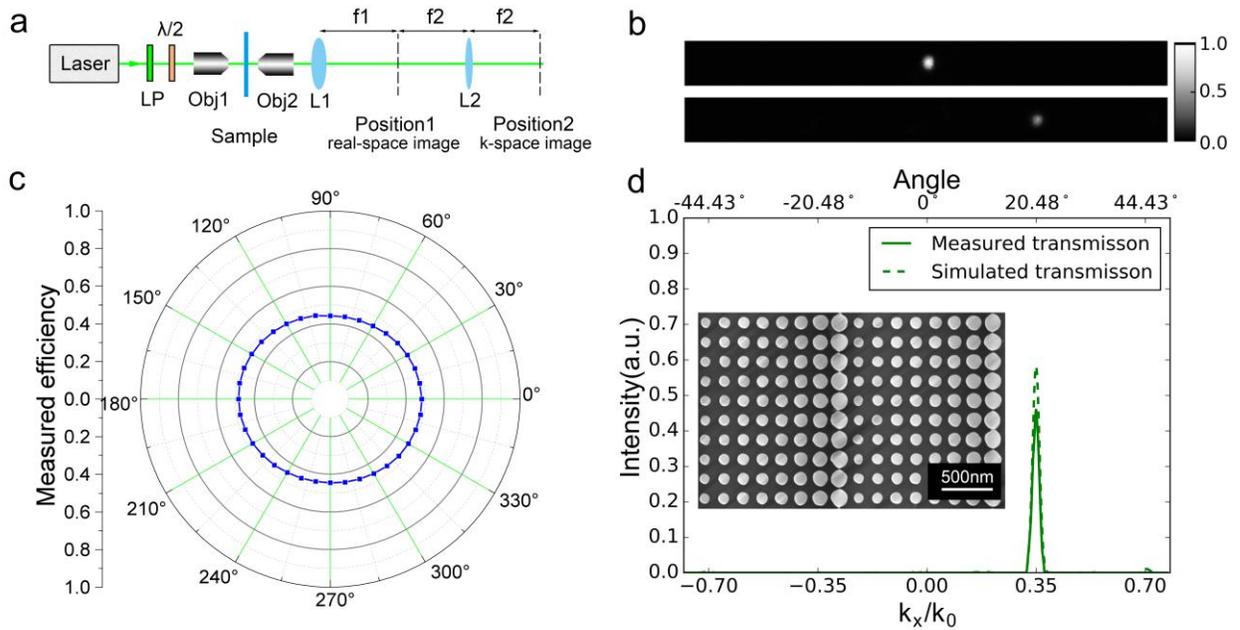

**Figure 6.** (a) Measurement set-up used to characterize the metasurfaces according to the design shown in Table 1. (b) Far-field profiles of the incident light intensity (top) and transmitted intensity (bottom) captured by a CCD camera. (c) Measured deflection efficiency by using equation (3) with different polarization directions. The polarization direction is defined as the angle between the electric field and gradient of the posts. (d) Experiment (solid line) and simulation (dashed line, same as Figure 3c) of the transmitted deflected beam intensity normalized to the input signal in $k_x$ direction. The inset is the SEM of the structure.

It is known that the transmission efficiency of plasmonic metasurfaces is limited to 25%[20] because of ohmic loss. Poly-silicon and a-silicon metasurfaces are limited in the shorter wavelength range because of absorption loss, and reported efficiencies are below 30% at 500nm even for polarization-dependent designs.[19] Low-index-contrast metasurfaces such as $TiO_2$[30, 31] offer higher transmission in the visible regime, but they also require higher thickness and very high aspect ratios to achieve a 0 to $2\pi$ phase shift for polarization-independent operation. By comparison and shown in Table 2, c-silicon metasurfaces offer significant advantages compared to these materials. Metasurfaces based on c-Si operate with a thinner film and a lower aspect ratio than $TiO_2$ and they achieve better transmission than poly-silicon and a-silicon in the visible regime. Based on a simple transfer technique, the c-silicon can be easily transferred to the desired substrate from the SOI wafer. Due to its high refractive index, the c-silicon pattern can be easily fabricated and surrounded with other low index materials to increase the numerical aperture of metalenses and flexible metasurfaces.[35, 40]

Having now experimentally demonstrated an efficiency of 47%, let us consider further improvements. First of all, we note a discrepancy of 12% between the simulated efficiency (59%, Figure 3c) and the experimental value. This discrepancy can be explained by fabrication tolerances; by analyzing the as-fabricated structures, we note an average size discrepancy in pillar diameter of 7nm. If we use this adjusted size in our simulation, the calculated efficiency becomes 45%, i.e. the same as the experimental value within measurement error. It is therefore realistic to achieve the predicted 59% value with a reduction in fabrication tolerances. Furthermore, we note that an increase in film thickness allows us to push the efficiency even further, i.e. up to 71% (Figure 4b). Demonstrating the possibility of achieving such a high efficiency for visible light with silicon is a truly surprising outcome of this work. These values are compared and put into context in Table 2.

**Table 2.** Summary of previously reported experimental metasurfaces operating in transmission in the near-infrared and the visible range. In the table, the deflection efficiency is defined as the ratio between the main diffraction order and the input power, the aspect ratio being defined a ratio of the minimum width of the nanostructure to the thickness of the material.

| Reference | Materials | Wavelength | Efficiency | Polarization | Building block | Thickness | Aspect ratio |
|---|---|---|---|---|---|---|---|
| Chen et al. [33] | Quartz | 633nm | 55% | Independence | Square posts | 1.38μm | 10 |
| Astilean et al. [41] | $TiO_2$ | 633nm | 83% | Linear | 1D grating | 540nm | 5.3 |
| Lalanne et al. [30] | $TiO_2$ | 633nm | 78% | Independence | Square posts | 487nm | 4.6 |
| Shalaev et al. [24] | a-silicon | 1550nm | 36% | Linear | Rectangular posts | 270nm | 1 |
| Yu et al. [12] | a-silicon | 705nm | 45% | Independence | Circular posts | 130nm | 0.93 |
| Lin et al.[19] | p-silicon | 500nm | 28% | Circular | Nano-beams | 100nm | 1.2 |
| This work (experiment) | c-silicon | 532nm | 47% | Independence | Circular posts | 220nm | 2.47 |
| This work (simulation) | c-silicon | 532nm | 59% | Independence | Circular posts | 220nm | 2.7 |
| This work (simulation) | c-silicon | 532nm | 71% | Independence | Circular posts | 250nm | 3.4 |

In summary, we have transferred thin film c-silicon onto a quartz substrate by adhesive wafer bonding, then demonstrated a c-silicon gradient metasurface for beam deflection at the wavelength of 532nm. Furthermore, our experiment demonstrates full 2π phase control. We demonstrate a polarization-independent deflection efficiency in transmission of 47% and show that this efficiency can be increased up to 71% in simulation, which is almost close to the values achieved with $TiO_2$, yet with much lower aspect ratio hence reduced fabrication complexity.

We believe that this approach can not only be applied to other wavefront shaping situations, such as focusing, vortex generation and holography, but also offers a viable route to efficient tunable metasurfaces on flexible substrates in the visible range. Our geometry is also attractive for a variety of applications in integrated optics, such as imaging, biomedical sciences or wearable consumer electronics.

AUTHOR INFORMATION

**Corresponding Author**

*E-mail: wangxueh@mail.sysu.edu.cn

**Author Contributions**

∥Z. Zhou and J. Li contributed equally to this work.

**Notes**

The authors declare no competing financial interest.

ACKNOWLEDGMENT

This work is supported by Ministry of Science and Technology of China (2016YFA0301300), Guangzhou science and technology projects (201607010044, 201607020023), Natural Science Foundation of Guangdong (2016A030312012), National Natural Science Foundation of China (11334015, 11304102), the Fundamental Research Funds for the Central Universities, the Open research project of the State Key Laboratory of Optoelectronic Materials and Technologies in Sun Yat-Sen University of China, and EPSRC of U.K. under Grant EP/J01771X/1 (Structured Light). We also would like to acknowledge Prof Jianwen Dong for useful discussions on the metasurfaces design.